 \documentclass[11pt,preprint ]{aastex} 

\shorttitle{X-ray Spectra of VY Scl Stars}
\shortauthors{Mauche \& Mukai}

 \slugcomment{Accepted for publication in 
             {\it Ap.J.\/} January 4, 2002}

\newcommand\Mwd   {M_{\rm wd}}
\newcommand\Rwd   {R_{\rm wd}}
\newcommand\Msun  {{\rm M_{\odot}}}
\newcommand\Mdot  {\dot{M}}
\newcommand\lax{{\lower0.75ex\hbox{ $<$ }\atop\raise0.5ex\hbox{ $\sim$ }}}
\newcommand\gax{{\lower0.75ex\hbox{ $>$ }\atop\raise0.5ex\hbox{ $\sim$ }}}

\begin{document}

\title{The X-ray Spectra of VY~Scl Stars Are Not Blackbodies}

\author{Christopher W.\ Mauche}
\affil{Lawrence Livermore National Laboratory, 
       L-43, 7000 East Avenue, Livermore, CA 94550}
\email{mauche@cygnus.llnl.gov}
\and
\author{Koji Mukai\altaffilmark{1}}
\affil{NASA/Goddard Space Flight Center, 
       Code 662, Greenbelt, MD 20771}
\email{mukai@milkyway.gsfc.nasa.gov}

\altaffiltext{1}{Also Universities Space Research Association}



\begin{abstract}

Using {\it ASCA\/} data, we find, contrary to other researchers using
{\it ROSAT\/} data, that the X-ray spectra of the VY~Scl stars TT~Ari and
KR~Aur are poorly fit by an absorbed blackbody model but are well fit by
an absorbed thermal plasma model. The different conclusions about the
nature of the X-ray spectrum of KR~Aur may be due to differences in the
accretion rate, since this star was in a high optical state during the
{\it ROSAT\/} observation, but in an intermediate optical state during
the {\it ASCA\/} observation. TT~Ari, on the other hand, was in a high
optical state during both observations, so directly contradicts the
hypothesis that the X-ray spectra of VY~Scl stars in their high optical
states are blackbodies. Instead, based on theoretical expectations and
the {\it ASCA\/}, {\it Chandra\/}, and {\it XMM\/} spectra of other
nonmagnetic cataclysmic variables, we believe that the X-ray spectra of
VY~Scl stars in their low and high optical states are due to hot thermal
plasma in the boundary layer between the accretion disk and the surface
of the white dwarf, and appeal to the acquisition of {\it Chandra\/} and
{\it XMM\/} grating spectra to test this prediction.

\end{abstract}


\clearpage 
 

\section{Introduction}

Cataclysmic variables (CVs) are a diverse class of semidetached binaries
including novae, dwarf novae, and novalike variables, composed typically
of a low-mass main-sequence secondary and a white dwarf. With the
exception of novae in outburst, the engine for all CVs is the release of
gravitational potential energy as material accretes onto the white dwarf.
In nonmagnetic systems accretion is mediated by a disk, and simple theory
predicts that half of the gravitational potential energy of the accreting
material is liberated in the disk and half is liberated in the boundary
layer between the disk and the surface of the white dwarf, with
luminosities $L_{\rm disk} \approx L_{\rm bl} \approx G\Mwd\Mdot /2\Rwd
=4\times 10^{34}  (\Mdot /10^{-8}~\Msun~{\rm yr^{-1}})(\Mwd/\Msun )
(\Rwd/10^9~{\rm cm})^{-1}~\rm erg~s^{-1}$, where $\Mdot $ is the
mass-accretion rate and $\Mwd $ and $\Rwd $ are respectively the mass and
radius of the white dwarf. When $\Mdot $ is low (e.g., dwarf novae in
quiescence), the boundary layer is optically thin and quite hot (of order
the virial temperature $T_{\rm vir}=G\Mwd m_{\rm H}/6k\Rwd\sim 20$ keV);
when $\Mdot $ is high (e.g., novalike variables and dwarf novae in
outburst), the boundary layer is optically thick and quite cool (of order
the blackbody temperature $T_{\rm bb}=[G\Mwd\Mdot /8\pi\sigma\Rwd
^3]^{1/4} \sim 10$ eV). In the context of this theory, high-$\Mdot $ CVs
are X-ray sources only because the upper ``atmosphere'' of the boundary
layer remains optically thin.

The observational picture of the  X-ray emission of CVs was first sketched
in the 1970s and has become clearer through the years as instruments with
higher effective area, broader bandpass, and better spectral resolution
have flown on X-ray satellites. Summaries of {\it HEAO~1\/}, {\it
Einstein\/}, {\it EXOSAT\/}, and {\it ROSAT\/} investigations are provided
by \citet{cor81, cor84, era91, muk93, ric96, tee96}; and \citet{ver97}.
That the X-ray emission region of nonmagnetic CVs is compact and centered
on the white dwarf is established directly by the X-ray light curves
of eclipsing systems. That the X-ray spectra of CVs are due to hot thermal
plasma was established by {\it ASCA\/}, which had the combination of large
effective area, broad bandpass, and high spectral resolution needed to
resolve the emission lines of K-shell Mg, Si, S, Ar, and Fe at high
energies, and to establish the presence of emission lines of K-shell O
and Ne and L-shell Fe at low energies \citep{nou94, muk00, bas01}. As
evidenced by {\it EUVE\/} light curves and spectra of the dwarf novae
SS~Cyg, U~Gem, VY~Hyi, and OY~Car in outburst, the blackbody component of
high-$\Mdot $ nonmagnetic CVs is typically not observed in the canonical
X-ray bandpass (SS~Cyg in outburst is the only clear exception, see
\citealt{beu93, pon95, mau95}) because its temperature is too low and
its luminosity is sometimes anomalously weak \citep[][and references
therein]{mau02}. While it is not yet possible to be entirely certain, the
working hypothesis is that the X-ray spectra of {\it all\/} nonmagnetic
CVs are due to hot thermal plasma in the boundary layer between the
accretion disk and the surface of the white dwarf. 

\section{VY Scl Stars}

Given this understanding, it is unsettling that the X-ray spectra of
VY~Scl stars have been repeatedly described as absorbed blackbodies.
VY~Scl stars are a class of novalike variables that occasionally {\it
dim\/} by 3--5 mag in the optical; they are sometimes referred to as
``anti-dwarf novae.'' It is thought that the downward transitions of
VY~Scl stars are the result of temporary reductions in the mass-transfer
rate from the secondary, possibly due to the passage of star spots over
the $L_1$ point. VY Scl stars are more than oddities in the CV zoo.
\citet{liv94} have proposed that the same mechanism that causes the
intensity dips in VY Scl stars (with $P_{\rm orb}=3$--4 hr) ultimately
results in the $P_{\rm orb}=2$--3 hr gap in the orbital period
distribution of CVs. \citet{lea99} have proposed that the lack of dwarf
nova outbursts in VY Scl stars in quiescence is the result of irradiation
of the inner disk by the hot ($T\gax 40$ kK) white dwarfs known to exist
in these systems. The white dwarfs in novalike variables should be
similarly hot because of their relatively high average mass-accretion
rates. Otherwise, novalike variables, VY~Scl stars in their high states,
and dwarf novae in outburst should all be quite similar. In particular,
we expect that their X-ray spectra are produced by hot thermal plasma
in the boundary layer between the accretion disk and the surface of the
white dwarf. 

The first indication that the X-ray spectra of VY~Scl stars might differ
from other high-$\Mdot $ CVs was provided by \citet{sch95}, who found
that the {\it ROSAT\/} PSPC spectra of the VY~Scl stars MV~Lyr and
KR~Aur are best described by absorbed blackbodies. Next, \citet{tee96}
published a figure showing that the {\it ROSAT\/} PSPC colors of the
VY~Scl stars TT~Ari, KR~Aur, BZ~Cam, and MV~Lyr lay in a region of
parameter space inhabited by blackbodies (and highly absorbed thermal
plasmas). Finally, \citet{gre98} concluded from a comprehensive study
of {\it ROSAT\/} pointed and all-sky survey PSPC observations that
the X-ray spectra of VY~Scl stars in their high optical states are best
described by absorbed blackbodies. Greiner fit the {\it ROSAT\/} PSPC
spectrum of the brightest VY~Scl star, TT~Ari, with absorbed power law,
thermal bremsstrahlung, Raymond-Smith thermal plasma, and blackbody
models and obtained $\chi_\nu^2=2.23$, 1.99, 1.87, and 1.67, respectively.
Although noting that even the absorbed blackbody model fit was poor,
Greiner advocated the blackbody interpretation of the {\it ROSAT\/}
PSPC spectra of VY~Scl stars as a class because it resulted in similar
parameters for the brightest members of the class: blackbody temperatures
$kT_{\rm bb}\approx 0.25$--0.5 keV and emitting sizes of 50--120~m (hence
fractional emitting areas $f= {\rm Area}/ 4\pi\Rwd ^2\lax 1\times
10^{-11}$).

There are a number of problems with the conclusion that the X-ray spectra
of VY~Scl stars are blackbodies. First, it runs counter to our theoretical
expectation that the X-ray emission of nonmagnetic CVs is due to hot
thermal plasma. Second, the implied emitting areas are far smaller than
any structure known in CVs. The area of the boundary layer $A_{\rm bl}
\approx 2\pi\Rwd\, 2H_{\rm bl}$, where $H_{\rm bl}$ is the vertical
height of the boundary layer. As argued by \citet{pat85}, a lower limit
to $H_{\rm bl}$ is the density scale height of the inner disk $H_{\rm 
disk}= 3\times 10^6\, (\Mdot /10^{-8}~{\rm \Msun~yr^{-1}})^{0.18} (\Mwd
/\Msun )^{1.2}$ cm \citep{pri79}. Hence, the boundary layer fractional
emitting area $f\gax H_{\rm disk}/\Rwd = 3\times 10^{-3}\, (\Mdot
/10^{-8}~{\rm \Msun~yr^{-1}})^{0.18} (\Mwd /\Msun )^{1.2} (\Rwd/10^9{~\rm
cm})^{-1}$. It is theoretically possible to produce even smaller
accretion spots by magnetically channeling the flow of material down
to the white dwarf surface, but even in CVs with the highest magnetic
field strengths (AM~Her stars or polars, with $B\sim 10$--100 MG), the
fractional emitting area $f\sim 10^{-3}$ \citep{mau99}. If the spot
sizes were this small in VY~Scl stars, it would be hard to avoid flux
modulations at the white dwarf spin period. Third, the blackbody
temperatures are too high for a white dwarf accretor. With $kT_{\rm bb}
\approx 0.25$--0.5 keV, the implied local energy flux $\sigma T_{\rm
bb}^4 = 4$--$64\times 10^{21}~\rm erg~cm^{-2}~s^{-1}$, orders of
magnitude greater than the Eddington rate: $L_{\rm edd}/4\pi\Rwd ^2
=1\times 10^{19}\, (\Mwd /\Msun ) (\Rwd/10^9~{\rm cm})^{-2}~\rm
erg~cm^{-2}~s^{-1}$.

\section{ASCA Observations}

Given the apparent discrepancy between {\it ROSAT\/} observations and
theory, we searched the {\it ASCA\/} archive for data to test the
hypothesis that the X-ray spectra of VY~Scl stars in their high optical
states are blackbodies. We found two VY~Scl stars in the archive: TT~Ari
and KR~Aur, observed by {\it ASCA\/} on 1994 January 20/21 and 1996 
March 6, respectively. By examining the AAVSO records, we determined
that on the dates of the {\it ASCA\/} observations TT~Ari was in a high
optical state with $V\approx 10.5$, while KR~Aur was in an intermediate
brightness state with $V\approx 16$ (whereas in its high and low states
$V\approx 14$ and $V\approx 19$, respectively). Therefore, the only
high-state VY~Scl star with data in the {\it ASCA\/} archive is TT~Ari,
but this is the brightest VY~Scl star, so provides the most stringent
test of the nature of the X-ray spectra of these high-$\Mdot $
nonmagnetic CVs. For completeness, we also analyzed the {\it ASCA\/}
spectrum of KR~Aur.

{\it ASCA\/} carries four co-aligned X-ray telescopes, two with gas
scintillation imaging proportional counter (Gas Imaging Spectrometer;
GIS) detectors and two with CCD (Solid-state Imaging Spectrometer; SIS)
detectors \citep{tan94}. During both observations, all four instruments
were operating normally, with the GIS detectors in standard PH mode. We
screened the GIS data by excluding intervals during Earth occultations
(up to $5^\circ $ above the horizon), passages through the South Atlantic 
Anomaly (SAA), and times of high background as indicated by the in-situ
monitors. After screening, we obtained 18.5~ks and 21.6~ks of good GIS
data for TT~Ari and KR~Aur, respectively. During the TT~Ari observation,
the SIS detectors were operated in 2-CCD clocking mode, switching between
Faint and Bright data mode depending on the available telemetry rate. We
applied a similar set of screening criteria (only differing in the
details of the monitor counts used) to the SIS data, and further excluded
data taken within $20^\circ $ of the bright Earth limb and 64~s of
day/night and SAA transitions. After the screening, we obtained 17.5~ks
of good SIS data for TT~Ari. During the KR~Aur observation, the SIS
detectors were used in 1-CCD Faint mode throughout, and after a similar
screening (but with $15^\circ $ limit for the bright Earth limb and 32~s
for day/night and SAA transitions), we obtained 21.1~ks of good SIS data.

Next, we extracted source and background spectra for each detector for
each observation. For TT~Ari, source counts were extracted from circular
regions centered on the source ($6'$ radius for the GIS, $2.5'$--$3.5'$
radius for the SIS to stay within a single CCD chip), while background
counts were extracted from adjoining source-free regions. Similar
extraction regions could not be used for KR~Aur because of the presence
of faint sources near the VY~Scl star. We therefore reduced the GIS
source extraction region to $4'$ radius, and used a local background
region that avoids both KR~Aur and the contaminating sources. For the
SIS, we used a similar source extraction region as for TT~Ari, but
extracted background spectra from blank-sky observations, since a
source-free local background region of sufficient size could not be
defined. For the spectral responses of the instruments, we have taken
GIS redistribution matrix v4.0, and generated SIS redistribution matrix
and auxiliary response files using the standard (HEAsoft v5.1) software.
For each observation, we co-added spectra and responses for the two GIS
instruments into one set, and the two SIS instruments into another set.
Finally, the SIS and GIS data sets were fit simultaneously after ignoring
bad channels, grouping the channels to produce a minimum of 20 counts
per channel, and restricting the energy range to 0.5--10 keV. Each data
set was fit with three spectral models: a single-temperature blackbody,
a single-temperature solar-abundance MEKAL thermal plasma, and a
two-temperature solar-abundance MEKAL thermal plasma, all absorbed by
a column density of neutral material. To account for known residual
discrepancies between the SIS and GIS response functions, we allowed
the absorbing column density to differ between the SIS and GIS.

The best-fit parameters (and 90\% confidence intervals for the thermal
plasma models) resulting from these fits are shown in Table~1. The 
single-temperature blackbody model produces very poor fits ($\chi_\nu^2
=5.77$ and 2.24) to the 0.5--10 keV {\it ASCA\/} spectra of TT~Ari and
KR~Aur because a blackbody produces too few counts both at high and low
energies. Even worse fits are realized with the blackbody parameters
$kT_{\rm bb}\approx 0.3$--0.5 keV and $N_{\rm H}\approx 1$--$2\times
10^{20}~\rm cm^{-2}$ favored by fits to the {\it ROSAT\/} PSPC spectra
of these stars. Far better fits ($\chi_\nu^2 =1.31$ and 0.75) are
obtained {\it with the same number of degrees of freedom\/} with the
single-temperature thermal plasma model. In particular, the Fe K$\alpha$
line is clearly detected near 6.7 keV in the GIS data of both sources,
which is decisive evidence for the presence of a hot ($kT\sim 5$--10 keV)
plasma. With two additional degrees of freedom, the two-temperature thermal
plasma model further improves the fits---by a significant amount in TT~Ari
($\Delta\chi^2=116$, $\chi_\nu^2=1.07$), and by a modest amount in KR~Aur
($\Delta\chi^2= 22$, $\chi_\nu^2=0.66$)---by better accounting for
systematic residuals at low energies. The best-fit two-temperature thermal
plasma model is shown in Figure~1 superposed on the SIS and GIS spectra
of TT~Ari.

\section{Summary}

Based on an analysis of {\it ASCA\/} SIS and GIS spectra, we find that
the 0.5--10 keV X-ray spectra of TT~Ari and KR~Aur are poorly fit by an
absorbed blackbody model but are well fit by an absorbed thermal plasma
model. The TT~Ari spectra are adequately described by a two-temperature
solar-abundance thermal plasma, with one component at $kT\approx 7$~keV
and another at $kT\approx 0.7$ keV, with a relative emission measure of
approximately 20:1 in favor of the high-temperature component. The
lower-quality KR~Aur spectra are adequately described by a
single-temperature solar-abundance thermal plasma with $kT\approx 6$~keV,
and allow a second temperature component only if the column density is
significantly increased and the emission measures of the two components
are comparable. These results should be understood to be simply
{\it parameterizations\/} of the X-ray spectra of these two stars, not
a definitive determination of the nature of their X-ray spectra.
Higher-quality data (e.g., {\it Chandra\/} High-Energy Transmission
Grating and {\it XMM\/} Reflection Grating Spectrometer and European
Photon Imaging Camera spectra) are required to determine---by resolving
the L-shell emission lines of Fe and the K-shell emission lines of H-
and He-like ions of abundant elements from C to Fe---the true emission
measure distribution, abundances, density, and Doppler broadening of
the plasma in these stars. It is clear, however, that {\it the X-ray
spectra of neither of these stars is that of a blackbody\/}, contrary
to the conclusions of \citet{sch95} and \citet{gre98}. The different
conclusions about the nature of the X-ray spectrum of KR~Aur may be due to
differences in the accretion rate, since this star was in a high optical
state during the {\it ROSAT\/} observation, but in an intermediate
optical state during the {\it ASCA\/} observation. TT~Ari, on the other
hand, was in a high optical state during both observations, so directly
contradicts the claim that the X-ray spectra of VY~Scl stars in their
high optical states are blackbodies. It remains possible that TT~Ari,
the brightest VY~Scl star, is anomalous, and that the X-ray spectra of
other high-state VY~Scl stars are blackbodies, but based on theoretical
expectations and the {\it ASCA\/}, {\it Chandra\/}, and {\it XMM\/} X-ray
spectra of other high-$\Mdot $ nonmagnetic CVs, we believe that they are
not. Instead, we believe that the X-ray spectra of VY~Scl stars in their
low and high optical states are due to hot thermal plasma in the boundary
layer between the accretion disk and the surface of the white dwarf, and
appeal to the acquisition of {\it Chandra\/} and {\it XMM\/} grating
spectra to test this prediction.

\acknowledgments

In this research, we have used data obtained from the High Energy
Astrophysics Science Archive Research Center (HEASARC), provided by NASA's
Goddard Space Flight Center.  We have also used, and acknowledge with
thanks, data from the AAVSO International Database, based on observations
submitted to the AAVSO by variable star observers worldwide. C.~W.~M.'s
contribution to this work was performed under the auspices of the U.S.\
Department of Energy by University of California Lawrence Livermore
National Laboratory under contract No.~W-7405-Eng-48.

\clearpage 


\clearpage 


\begin{deluxetable}{lccccccc}
\tablecolumns{8} 
\tablewidth{0pc} 
\tablenum{1}
\tablecaption{Model Parameters\label{tab1}}
\tablehead{
\colhead{}& \colhead{SIS $N_{\rm H}$}& \colhead{GIS $N_{\rm H}$}& \colhead{$kT_1$}& \colhead{}& \colhead{$kT_2$}& \colhead{}& \colhead{}\\
\colhead{Model}& \colhead{$(\rm cm^{-2})$}& \colhead{$(\rm cm^{-2})$}& \colhead{(keV)}& \colhead{$N_1$\tablenotemark{a}}& \colhead{(keV)}& \colhead{$N_2$\tablenotemark{a}}& \colhead{$\chi^2/\nu $}\\
}
\startdata 
\cutinhead{TT Ari} 
Blackbody&  0                       &  0                       &  0.70              & $    1.3              \, $E31& \nodata               & \nodata                  & $   2752/477=5.77$\\
1T MEKAL & $6.8^{+1.0}_{-1.0}\, $E20& $0.0^{+9.9}_{-0.0}\, $E19& $7.1^{+0.3}_{-0.4}$& $   11.1^{+0.2}_{-0.2}\, $E53& \nodata               & \nodata                  & $\phn623/477=1.31$\\
2T MEKAL & $1.4^{+0.4}_{-0.3}\, $E21& $7.7^{+3.8}_{-4.5}\, $E20& $7.4^{+1.0}_{-0.4}$& $   11.0^{+0.2}_{-0.3}\, $E53& $0.68^{+0.14}_{-0.05}$& $5.9^{+1.8}_{-1.4}\, $E52& $\phn507/475=1.07$\\
\cutinhead{KR Aur} 
Blackbody&  0                       &  0                       &  0.61              & $    2.5              \, $E30& \nodata               & \nodata                  & $\phn493/220=2.24$\\
1T MEKAL & $7.6^{+3.5}_{-2.9}\, $E20& $0.0^{+6.2}_{-0.0}\, $E20& $5.8^{+1.0}_{-0.7}$& $\phn2.3^{+0.1}_{-0.1}\, $E53& \nodata               & \nodata                  & $\phn165/220=0.75$\\
2T MEKAL & $8.1^{+1.2}_{-7.8}\, $E21& $7.6^{+1.3}_{-5.9}\, $E21& $6.8^{+3.4}_{-2.1}$& $\phn2.3^{+0.3}_{-0.2}\, $E53& $0.55^{+0.29}_{-0.08}$& $2.7^{+1.3}_{-1.9}\, $E53& $\phn143/218=0.66$\\
\enddata
\tablenotetext{a}{Normalization expressed as luminosity $(\rm erg~s^{-1})$ for the blackbody model
and  emission measure $(\rm cm^{-3})$ for the MEKAL model, both for a fiducial distance of 100 pc.}
\end{deluxetable}

\clearpage 


\begin{figure}
\figurenum{1}
\epsscale{0.932}
\plotone{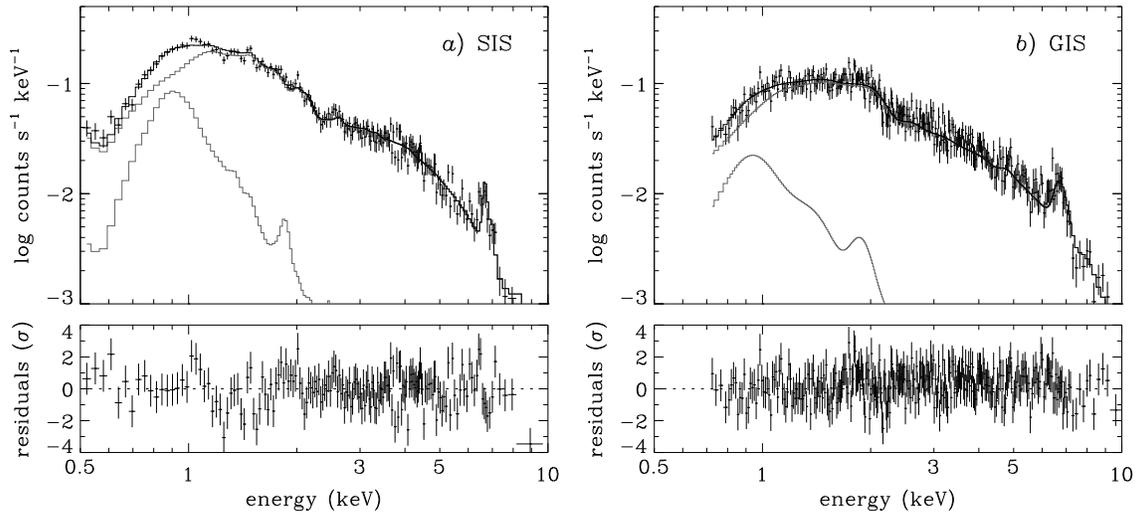}
\caption{{\it ASCA\/} SIS and GIS spectra, best-fit two-temperature
thermal plasma models, and residuals for TT~Ari.}
\end{figure}


\begin{thebibliography}{} 

\bibitem[Baskill, Wheatley, \& Osborne(2001)]{bas01}
         Baskill, D.~S., Wheatley, P., \& Osborne, J.
         2001, \mnras , 328, 71
\bibitem[Beuermann \& Thomas(1993)]{beu93}
         Beuermann, K., \& Thomas, H.-C. 1993, Adv.\ Space Sci., 13, 115
\bibitem[C\'ordova, Jensen, \& Nugent(1981)]{cor81}
         C\'ordova, F.~A., Jensen, K.~A., \& Nugent, J.~J.
         1981, \mnras , 196, 1
\bibitem[C\'ordova \& Mason(1984)]{cor84}
         C\'ordova, F.~A., \& Mason, K.~O. 1984, \mnras , 206, 879
\bibitem[Eracleous, Halpern, \& Patterson(1991)]{era91}
         Eracleous, M., Halpern, J., \& Patterson, J. 1991, \apj , 
         382, 290
\bibitem[Greiner(1998)]{gre98}
         Greiner, J. 1998, \aap , 336, 626
\bibitem[Leach et al.(1999)]{lea99}
         Leach, R., Hessman, F.~V., King, A.~R., Stehle, R., \& Mattei,
         J. 1999, \mnras , 305, 225.
\bibitem[Livio \& Pringle(1994)]{liv94}
         Livio, M., \& Pringle, J.~E. 1994, \apj , 427, 956
\bibitem[Mauche(1999)]{mau99}
         Mauche, C.~W. 1999, in Annapolis Workshop on Magnetic
         Cataclysmic Variables, ed.\ C.~Hellier \& K.~Mukai (San
         Francisco: ASP), 157
\bibitem[Mauche(2002)]{mau02}
         Mauche, C.~W. 2002, in Continuing the Challenge of EUV Astronomy:
         Current Analysis and Prospects for the Future, ed.\ S.~Howell,
         et al.\ (San Francisco: ASP), in press (astro-ph/0109133)
\bibitem[Mauche, Raymond, \& Mattei(1995)]{mau95}
         Mauche, C.~W., Raymond, J.~C., \& Mattei, J.~A. 1995, \apj ,
         446, 842
\bibitem[Mukai(2000)]{muk00}
         Mukai, K. 2000, New Astr.\ Rev., 44, 9
\bibitem[Mukai \& Shiokawa(1993)]{muk93}
         Mukai, K., \& Shiokawa, K. 1993, \apj , 418, 863
\bibitem[Nousek et al.(1994)]{nou94}
         Nousek, J.~A., et al. 1994, \apj , 436, 19
\bibitem[Patterson \& Raymond(1985)]{pat85}
         Patterson, J. \& Raymond, J.~C. 1985, \apj , 292, 550
\bibitem[Ponman et al.(1995)]{pon95}
         Ponman, T.~J., et al. 1995, \mnras , 276, 495
\bibitem[Pringle \& Savonije(1979)]{pri79}
         Pringle, J.~E., \& Savonije, G.~J. 1979, \mnras , 187, 777
\bibitem[Richman(1996)]{ric96}
         Richman, H.~R. 1996, \apj , 462, 404
\bibitem[Schlegel \& Singh(1995)]{sch95}
         Schlegel, E.~M., \& Singh, J. 1995, \mnras , 276, 1365
\bibitem[Tanaka, Inoue, \& Holt(1994)]{tan94}
         Tanaka, Y., Inoue, H., \& Holt, S.~S. 1994, \pasj , 46, L37
\bibitem[van Teeseling, Beuermann, \& Verbunt(1996)]{tee96}
         van Teeseling, A., Beuermann, K., \& Verbunt, F.
         1996, \aap , 315, 467
\bibitem[Verbunt et al.(1997)]{ver97}
         Verbunt, F., Bunk, W.~H., Ritter, H., \& Pfeffermann, E. 
         1997, \aap , 327, 614

\end{thebibliography}
\end{document}